\documentstyle[aps,prl,preprint,floats,epsfig]{revtex}

\begin{document}


\tighten

\long\def\figz&#1&#2{\begin{figure}[phtb] \vspace{10.4 cm}
   \includegraphics{#1}
   \caption[#2]{\label{fig:#1}{#2}} \end{figure}}

\preprint{\vbox{\hbox{\bf CLNS 98/1545   \hfill}
                \hbox{\bf CLEO 98-5   \hfill}
                \hbox{\bf \today       \hfill}}}

\title{\boldmath Measurement of the Mass Splittings between the
b\=b $\chi_{b,J}$(1P) States}

\author{CLEO Collaboration}
\date{\today}

\maketitle
\tighten

\begin{abstract}
We present new measurements of photon energies and branching
fractions for the radiative transitions: 
$\Upsilon$(2S)$\to\gamma\chi_{b(J=0,1,2)}$(1P).
The masses of the $\chi_b$ states are determined
from the measured radiative photon energies. 
The ratio of mass splittings between the
$\chi_b$ substates, $r\equiv(M_{J=2}-M_{J=1})/(M_{J=1}-M_{J=0})$,
with $M$ the $\chi_b$ mass, provides information on the nature of the
b\=b confining potential. We find
$r$(1P)=$0.54\pm0.02\pm0.02$. This value is in conflict
with the previous world average, but more consistent
with the theoretical expectation that $r$(1P)$<r$(2P); i.e., that this
mass splittings ratio is smaller for the $\chi_b$(1P) states than
for the $\chi_b$(2P) states.
\end{abstract}

\newpage

\begin{center}
K.~W.~Edwards,$^{1}$
A.~Bellerive,$^{2}$ R.~Janicek,$^{2}$ D.~B.~MacFarlane,$^{2}$
P.~M.~Patel,$^{2}$
A.~J.~Sadoff,$^{3}$
R.~Ammar,$^{4}$ P.~Baringer,$^{4}$ A.~Bean,$^{4}$
D.~Besson,$^{4}$
% ...
\end{center}
 
\small
\begin{center}
$^{1}${Carleton University, Ottawa, Ontario, Canada K1S 5B6 \\
and the Institute of Particle Physics, Canada}\\
$^{2}${McGill University, Montr\'eal, Qu\'ebec, Canada H3A 2T8 \\
and the Institute of Particle Physics, Canada}\\
$^{3}${Ithaca College, Ithaca, New York 14850}\\
$^{4}${University of Kansas, Lawrence, Kansas 66045}\\
% ...
$^{12}${Southern Methodist University, Dallas, Texas 75275}\\
$^{13}${Syracuse University, Syracuse, New York 13244}\\
% ...
\end{center}

\newpage
\section{ Introduction }

The $\Upsilon$ particles (bound systems of heavy bottom--anti-bottom quarks)
play an important role in studies of strong interactions.
The bottom--anti-bottom pair creates 
a positronium-like system bound by strong interactions. Because the 
$\Upsilon$ system is nearly non-relativistic ($\beta^2 \approx 0.08$), 
theory can start from a relatively simple non-relativistic 
potential model and add relativistic terms as next-order corrections 
to describe the $\Upsilon$ mass-spectrum, as well as the
partial widths for the
transitions expected within the bottomonium system. 
Relativistic effects as a result of 
spin-orbit and tensor interactions generate fine splittings; hyperfine 
splittings arise from spin-spin interactions.
Potential models
predict electric dipole transitions 
$\Upsilon$(2S)$\rightarrow \gamma\chi_{b,J}$(1P)
with rates proportional to $(2J+1)E^{3}_{\gamma}$, with $E_{\gamma}$
 the photon energy. These transitions have already been investigated
in four experiments -- 
CUSB\cite{cusb}, CLEO\cite{cleo}, Crystal Ball\cite{cball} 
and ARGUS\cite{argus} -- by measuring the energy distribution of
transition photons detected inclusively in multi-hadronic events:
$\Upsilon$(2S)$\rightarrow \gamma\chi_{b,J}$(1P);
$\chi_{b,J}$(1P)$\rightarrow $ hadrons. 
The exclusive radiative cascade
transitions, $\Upsilon$(2S)$\rightarrow \gamma\chi_{b,J}$(1P);
$\chi_{b,J}$(1P)$\rightarrow \gamma\Upsilon$(1S), 
in which the $\Upsilon$(1S) is tagged
by its decay to dileptons,
were measured by the CUSB\cite{cusb1} and
Crystal Ball\cite{cball1} 
experiments.

In the present analysis, we have used the inclusive method
to study the radiative photon transitions 
between the $\Upsilon$(2S) 
and the $\chi_{b,J}$(1P)
and measure the 
fine structure of the $P$ states,
which can be characterized by the ratio of mass
splittings within the triplet: 
$r\equiv(M_{J=2}-M_{J=1})/(M_{J=1}-M_{J=0})$. 
The ratio of mass splittings measured in $\Upsilon$(3S)
radiative transitions to the $\chi_{b,J}$(2P) triplet
is $r$(2P)=$0.58\pm0.01$\cite{PDG}. 
Phenomenologically,
the parameter $r$ gives information on the Lorentz transformation properties
of the b\=b confining potential; different predictions for $r$ result from
different assumptions about the relative vector and scalar contributions.
The tabulated world average
for the ratio of mass splittings 
measured for the $\chi_{b,J}$(1P) triplet is $r(1P)=0.65\pm0.03$\cite{rev}, 
corresponding to
$r(2P)<r(1P)$, opposite to most model predictions\cite{rev}. 

\section{Data Sample and Analysis Description}

These data were obtained with the CLEO II detector\cite{NIM} 
at the Cornell Electron
Storage Ring, CESR, 
corresponding to an
integrated luminosity of 73.6 pb$^{-1}$ at the
$\Upsilon$(2S) energy. 
Based on the number of hadronic events measured at this energy, we
determine that
this luminosity is equivalent to a total number of
(488$\pm$18)$\cdot10^{3}$ produced $\Upsilon$(2S) events.
The advantage of the present analysis over previous analyses lies primarily
in the high segmentation of the CLEO II calorimeter, which offers
improved resolution of individual photon showers, with excellent solid
angle coverage.

Candidate events are required to have at least three observed charged tracks
in the event, with a total visible energy greater than the single
electron (or positron) beam energy. Additional criteria are imposed to
minimize contamination to our photon spectrum from non-hadronic events,
such as beam-gas, beam-wall, or two-photon collisions\cite{chi2p}. 
We note that such backgrounds 
contribute only a smooth background to our
observed photon energy spectrum.

\subsection{Measurement of Photon Transition Energies}

Only photons from
the barrel region ($|\cos\theta_{\gamma}| < 0.7$, with $\theta_{\gamma}$ 
the polar angle of the shower) are considered in this
analysis. The 
fractional energy resolution for photons in the barrel region of
the calorimeter ($\sigma_E\approx$5\% for $E_\gamma$=130 MeV) is
approximately twice as good as in the endcap regions.
Photon candidates are
required to be well separated from charged tracks and other photon
candidates in the same event.
The lateral shower shape is required to be consistent with that expected from
a true photon, and inconsistent
with the energy deposition patterns expected for charged particles. 
Showers from
``hot spots'' in the calorimeter are flagged on a run-by-run basis
and eliminated from consideration as inclusive photon candidates.

The photon energy scale is set by a three-stage calorimeter calibration
procedure\cite{NIM,chi2p}.
Pulsing of the readout electronics enables
determination of the pedestals and gains characteristic of each channel, 
independent of the crystal light output. The energy calibrations for 
individual crystals are then calculated using reconstructed showers
matched to beam-energy electrons in Bhabha events; the factors which 
convert normalized crystal light output to energy deposition, one 
per crystal, are obtained by minimizing the r.m.s. width of the electron 
shower energy distribution and constraining it to peak at the beam
energy. The third and final stage of calibration guarantees that any 
monochromatic photon energy spectrum peaks at the incident photon
energy, effectively correcting for any non-linearity in crystal
response, shower leakage from the cesium iodide, or bias in the 
reconstruction algorithm. This absolute energy calibration selects
photon candidates that can be kinematically constrained, using radiative
Bhabha ($ee\gamma$), $\gamma\gamma$ ($\gamma\gamma\gamma$), and muon pair 
($\mu\mu\gamma$) events for photons above 0.5~GeV and $\pi^0$'s below
2.5~GeV. Most relevant to this analysis is the $\pi^0$ calibration, 
which requires consistency between the observed $\pi^0\to\gamma\gamma$
mass peak and the expected mass. The $\pi^0$ calibration accounts for
the contributions to the observed $\pi^0$ mass lineshape from
energy-dependent shower angle and energy resolutions of both its
constituent photons. The correction amounts to $\sim$1\% near 100~MeV,
and varies slowly and continuously with energy. The $ee\gamma$, 
$\mu\mu\gamma$, $\gamma\gamma\gamma$, and $\pi^0$ samples yield 
compatible corrections in the energy regions where they can be
compared with one another. For the energy regime in this analysis we 
assess the uncertainty in the overall absolute energy scale to be $\pm$0.4\%.

After applying all event selection requirements and photon criteria,
we obtain the inclusive photon spectrum 
shown in 
Figure \ref{fig:3350198-001.ps}, with a fit to 
signal plus background overlaid.
\figz&{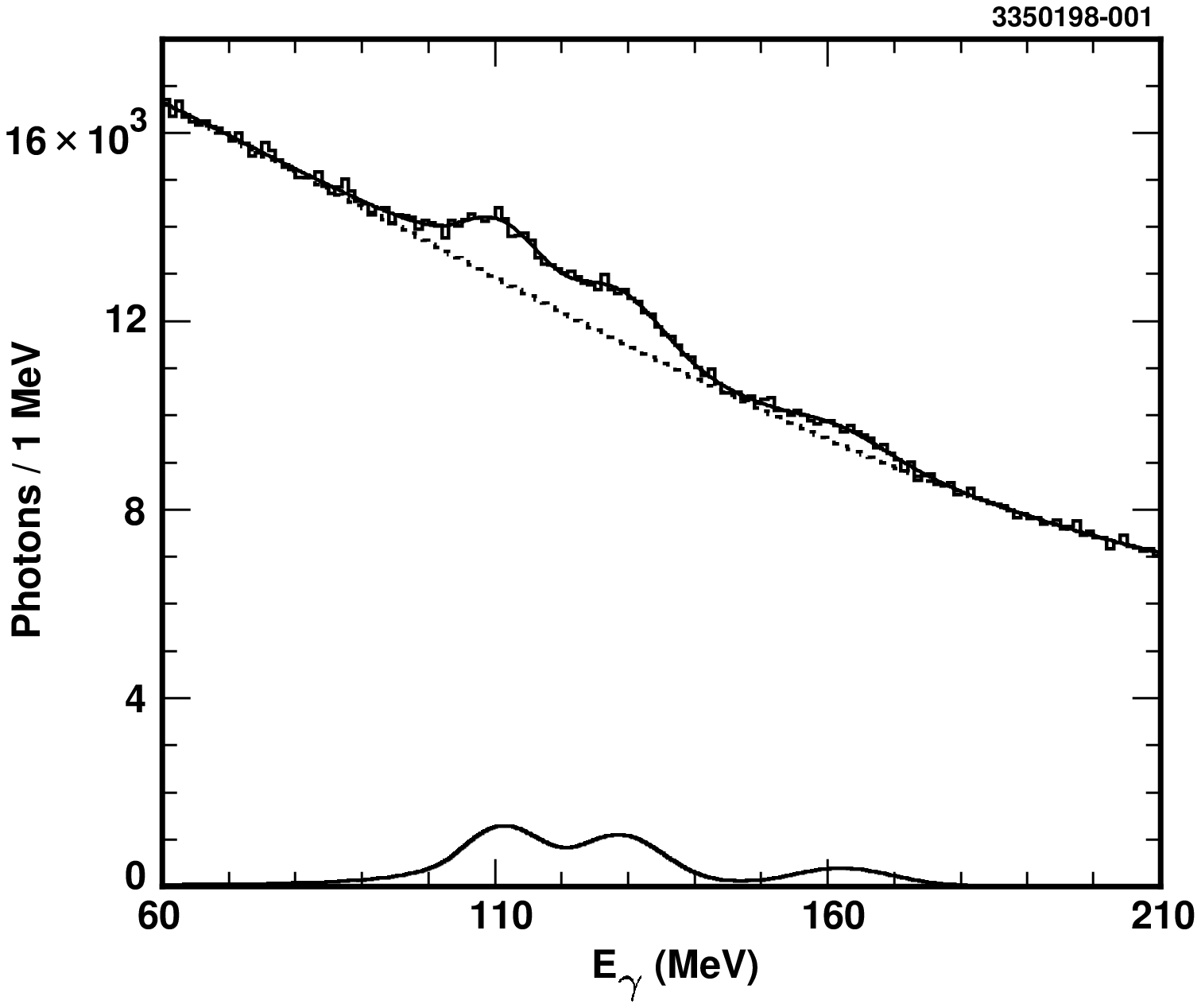}&{Fit 
to the inclusive spectrum
of photon candidates for the $\Upsilon$(2S) data set. Also shown is the
signal function used in the fit.}
Three enhancements are visible in this distribution above a smooth background.
We attribute the
lower energy photon peak to
$\Upsilon$(2S)$\rightarrow\gamma\chi_{b,2}$(1P),
the middle peak to $\Upsilon$(2S)$\rightarrow\gamma\chi_{b,1}$(1P), 
and the highest energy peak to the
$\Upsilon$(2S)$\rightarrow\gamma\chi_{b,0}$(1P) transition. 
The smooth
background is 
primarily due to 
$\pi^0 \rightarrow \gamma \gamma$ photons, 
as well as non-photon showers 
which passed the photon selection criteria. We fit this background shape
using a third order polynomial. 

The signal shape is parameterized using a functional form originally
used by the Crystal Ball collaboration\cite{TSthesis}. 
This is a nearly Gaussian
distribution with a tail at lower energies to take into account
longitudinal and transverse shower leakage in the calorimeter.
The expected ratios of the line widths (i.e., the shape of the
resolution curve as a function of photon energy) 
are fixed from Monte Carlo simulations; the width of
one of the lines (J=1) is allowed to float in the fit.
Since the intrinsic widths of the $\chi_{b,J}$(1P) states are expected to
be of order $\le$1 MeV, the experimental resolution should dominate
the observed line width.
The aggregate signal function therefore
has seven free parameters: the three line
positions, their areas,
and the energy resolution of the middle photon
line. 
The spectrum of photon 
candidates after the third order polynomial is subtracted 
is shown in 
Figure
\ref{fig:3350198-002.ps}a), with
the aggregate signal function overlaid. The results of 
the fit are given in Table \ref{tab:mnfit} (only statistical 
errors are presented). 
In Figure \ref{fig:3350198-002.ps}b),
we have superimposed upon the background-subtracted
spectrum the signals that would be expected, 
using the presently tabulated values for the masses of
the $\chi_{b,J}$(1P) states\cite{PDG}. It is clear that this results in
an inferior fit.
Our background-subtracted data are obviously incompatible
with the presently tabulated $\chi_{b,J}$(1P) masses.
From the fit in Fig. \ref{fig:3350198-002.ps}a), 
$r$(1P) is determined to be
$0.54\pm0.02$ (statistical error only). 

\figz&{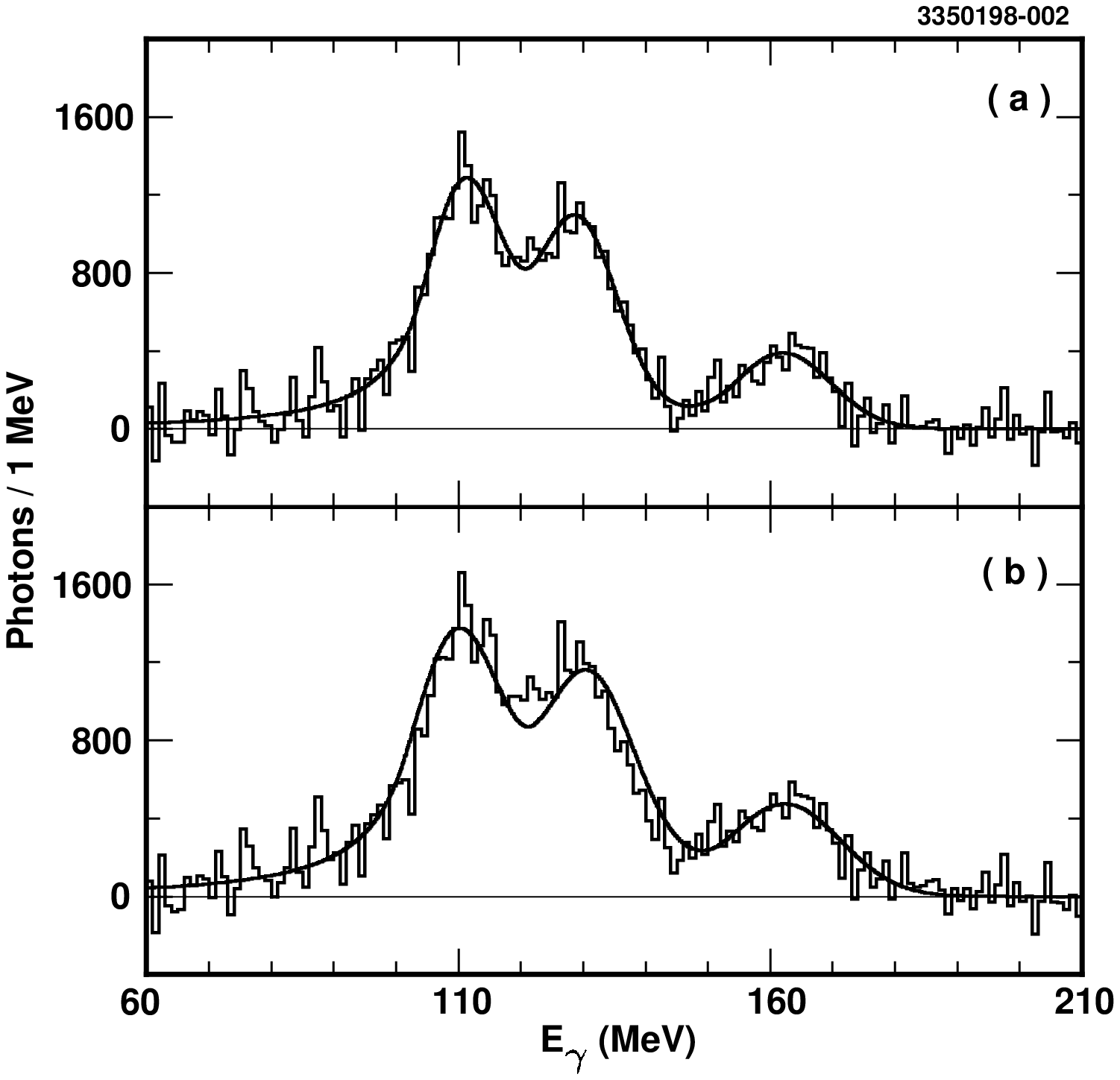}&{Background subtracted
photon energy spectrum, showing results of a free fit (top curve overlaying 
the histogram), and also overlay
of the signal function curve obtained by
constraining the masses of the $\chi_{b,J}$(1P) masses to their
previously tabulated values (bottom).}

\begin{table}[h]
\center
\caption[]{\small \label{tab:mnfit} Energies, raw yields (from
fit to data photon spectrum) and efficiencies
(from Monte Carlo simulations, and including J-dependent
geometric
acceptances) for $\Upsilon$(2S)$\to\gamma\chi_{b,J}$(1P)
transitions. Errors shown are statistical only.}
\begin{tabular}{cccc}
Transition & $E_{\gamma}$ (MeV) & Yield ($N_\gamma$) & Efficiency (\%) \\
\hline
 $J=2$ & $110.8\pm0.3$ & $20723\pm1436$ & $57.1\pm5.2$\\
 $J=1$ & $128.8\pm0.4$ & $20806\pm1466$ & $52.4\pm7.1$ \\
 $J=0$ & $162.0\pm0.8$ & $8637\pm1274$ & $61.8\pm5.9$ \\
\end{tabular}
\end{table}

\subsection{Branching Fractions}

We use a GEANT-based, full CLEO II Monte Carlo simulation to determine
the photon finding efficiency for each transition line, as presented
in Table \ref{tab:mnfit}.
The geometric acceptance is calculated for each transition using
the appropriate polar angular distributions
for transitions to 
the $J=2$ ($dN/dcos\theta_{\gamma}\propto 1+\frac{1}{13}cos^{2}\theta_\gamma$),
$J=1$ ($dN/dcos\theta_{\gamma}\propto 1+cos^{2}\theta_\gamma$),
and $J=0$ ($dN/dcos\theta_{\gamma}\propto 1-\frac{1}{3}cos^{2}\theta_\gamma$)
states, respectively. 
We use the
acceptance appropriate for
each transition (Table \ref{tab:mnfit})
to calculate the branching fractions from the $\Upsilon$(2S)
to the $\chi_{b,J}$(1P) triplet states (Table \ref{tab:concl}).


\subsection{Systematic Uncertainties and Cross-checks}

In addition to the systematic error due to the calorimeter calibration, 
primary systematic errors are due to
event and photon selection and the
signal extraction procedure, as summarized in
Table \ref{tab:tse}. The photon selection systematic is evaluated by 
remeasuring the photon spectrum using different definitions of `photons'.
We estimate the signal extraction systematic by using a variety of
signal parameterizations (using a bifurcated Gaussian rather than the 
Crystal Ball line shape, e.g.) and different parameterizations for the
background under the signal. The larger overall systematic error for
the $J=0$ line is attributable to the closer proximity of the minimum
ionizing peak for this line compared to the two lower energy lines
(and hence, greater sensitivity to photon selection requirements designed
to suppress showers from charged tracks) and a greater uncertainty in the
extrapolated energy resolution at this energy.
\begin{table}[h]
\center
\caption[]{\small \label{tab:tse} Evaluation of the total systematic errors.}
\begin{tabular}{cccccc}
Uncertainty in: & Calibration error & selection & fitting & number of   & total \\
          &             &           &         & $\Upsilon$(2S) produced & \\
\hline
$J=2$ line position (MeV) & 0.40  & 0.27 & 0.38 &-&   0.6\\
$J=1$ line position (MeV) & 0.42  & 0.26 & 0.31 &-&   0.6\\
$J=0$ line position (MeV) & 0.45  & 0.62 & 0.93 &-&   1.2\\
 $r$   & 0.001 &0.012 & 0.021&-&   0.02\\
${\cal B}(\Upsilon({\rm 2S})\to\gamma\chi_{b,2})$ 
 (\%)&-&4.6 & 5.9 &3.7& 8.3\\
${\cal B}(\Upsilon({\rm 2S})\to\gamma\chi_{b,1})$ 
(\%)&-&5.9 & 7.5 &3.7& 10.2\\
${\cal B}(\Upsilon({\rm 2S})\to\gamma\chi_{b,0})$ 
(\%)&-&10.3 &11.6 &3.7& 15.9\\ \vspace{0.2cm}
$\frac{|\langle\chi_{b,2}({\rm 1P})
|R|{\rm 2S}\rangle|^2}{|\langle\chi_{b,1}({\rm 1P})|R|{\rm 2S}\rangle|^2}$ &
-&0.055&0.096&-&0.11 \\ \vspace{0.2cm}
$\frac{|\langle\chi_{b,0}({\rm 1P})|R|{\rm 2S}\rangle|^2}
{|\langle\chi_{b,1}({\rm 1P})|R|{\rm 2S}\rangle|^2}$ &
-&0.041&0.255&-&0.26 \\
\end{tabular}
\end{table}

As a cross-check on the extracted value of $r$, we
have conducted a parallel analysis, in which we search for 
photons in the `exclusive' mode. In this case, we fully
reconstruct the decay chain:
$\Upsilon$(2S)$\to\chi_{b,J}$(1P)$\gamma$; 
$\chi_{b,J}$(1P)$\to\gamma\Upsilon$(1S);
$\Upsilon$(1S)$\to l^+l^-$, for which
$l^+l^-$ is $e^+e^-$ or $\mu^+\mu^-$. 
This very distinctive final state topology
consists of two leptons and two photons.
Unfortunately, due to the very small branching fraction for
$\chi_{b,0}$(1P)$\to\gamma\Upsilon$(1S), these exclusive events cannot be
used to completely
determine $r$(1P). Nevertheless, 
we find that the
the measured mass difference between the $J=2$ and $J=1$ states from
our exclusive data ($\Delta M=129.9\pm0.7-111.0\pm1.1$ MeV$=18.9\pm1.3$ MeV,
statistical errors only)
is in agreement with the mass difference measured in the 
inclusive mode ($\Delta M=18.0\pm1.0$ MeV, as computed from Table
\ref{tab:concl}).
We can also combine the masses observed for the $J=2$ and $J=1$ states in the
exclusive mode with the mass measured for the $J=0$ state in the
inclusive mode to obtain a value of $r'$; the superscript here
indicates that this quantity is derived from a combination of the
exclusive and the inclusive measurements. We obtain
$r'=0.59\pm0.05$ (statistical errors only), consistent with the
value we obtained from the inclusive data. 
We do not include these exclusive results in our 
final determination of $r$(1P) owing to their relatively small statistical
weight compared to the inclusive sample.

\subsection{Comparison with Previous Experimental Results}

Table \ref{tab:concl} summarizes the results for
the photon energies from our
inclusive analysis
and compares our results with the 
present Particle Data Group averages. We have also tabulated the
$\chi_{b,J}$(1P) masses inferred from our measured
photon energies.
Table \ref{tab:concl} similarly compares our results for the
branching fractions with previous measurements.
We find $r$(1P)
to be $0.54\pm0.02\pm0.02$,
inconsistent with the previous world average of $0.65\pm0.03$.
Note that $r$(1P) is insensitive to an overall miscalibration of the
photon energy scale.

\begin{table}[h]
\center
\caption[]{\small \label{tab:concl} 
$\Upsilon$(2S)$\to\gamma$X Transition energies and
branching fractions (${\cal B}$).}
\begin{tabular}{cccccc}
Transition& $E_{\gamma}$ (This Expt.) & 
$E_\gamma$ PDG\cite{PDG} & 
$M(\chi_{b,J})$(1P) (This Expt.) &
${\cal B}$ (This Expt.) & PDG\cite{PDG}     \\
          &  (MeV)           & (MeV)  & (MeV) & (\%) & (\%) \\
\hline
 $J=2$ & $110.8\pm0.3\pm0.6$ &  $109.6\pm0.6$ & $9912.5\pm0.3\pm0.6$ & 
$7.4\pm0.5\pm0.6$ & $6.6\pm0.9$\\
 $J=1$ & $128.8\pm0.4\pm0.6$ &  $130.6\pm0.7$& $9894.5\pm0.4\pm0.6$ & 
$6.9\pm0.5\pm0.7$ & $6.7\pm0.9$\\
 $J=0$ & $162.0\pm0.8\pm1.2$ &  $162.3\pm1.3$& $9863.0\pm0.8\pm1.2$ & 
$3.4\pm0.5\pm0.5$ & $4.3\pm1.0$\\
\end{tabular}
\end{table}

The widths for
the electric dipole transitions $\Upsilon$(2S)$\to\gamma\chi_{b,J}$(1P)
are given in terms of the characteristic interquark separation $R$ by:

\[\Gamma_{E1}={\cal B}\Gamma_{tot}=
\frac{4}{27}\alpha e^{2}_{b}E^{3}_{\gamma}(2J+1)
|\langle {\rm 1P}|R|{\rm 2S} \rangle|^{2}, \]

\noindent
In this equation, 
$\alpha$ is the electromagnetic coupling constant, $e_{b}$ is
the charge of b quark, and $\langle {\rm 1P}|R|{\rm 2S} \rangle$
is the matrix element. By averaging over the transitions to all three
$\chi_{b,J}$(1P) states and using $\Gamma_{tot}(2S)=(44.0\pm7.0)$ 
keV\cite{PDG}, we find $\langle {\rm 1P}|R|{\rm 2S}\rangle=
(1.88\pm0.12)$ GeV$^{-1}$,
in which the error includes both statistical and systematic errors.
This can be compared to the world average 
$\langle {\rm 1P}|R|{\rm 2S} \rangle=(1.9\pm0.2)$ GeV$^{-1}$ \cite{rev}.

By determining
the ratios of the
transition widths for states having 
different total angular momentum $J$, we can cancel the 
uncertainty due to the total $\Upsilon$(2S) width. Ratios of
the squared matrix element for different $J$ values are equal
to ratios of the quantity $\Gamma_{E1}/(E^{3}_{\gamma}(2J+1))$.
Table \ref{tab:concl-rat1} presents our experimental results and the
previous
world average for ratios of $\Gamma_{E1}/(E^{3}_{\gamma}(2J+1))$.
Theoretically these ratios are expected to be 1.0
     in the non-relativistic limit. Spin dependence
     of the matrix element 
     is introduced by relativistic corrections. Although
calculations vary, all models predict that the
J-dependent corrections follow:
$|\langle \chi_{b,2}({\rm 1P})|R|{\rm 2S}\rangle |^2>
|\langle \chi_{b,1}({\rm 1P})|R|{\rm 2S}\rangle |^2>
|\langle \chi_{b,0}({\rm 1P})|R|{\rm 2S}\rangle |^2$\cite{rev}. 

\begin{table}[h]
\center
\caption[]{\small \label{tab:concl-rat1} Ratios of 
$\Gamma_{E1}/(E^{3}_{\gamma}(2J+1))$.}
\begin{tabular}{ccc}
   & $\frac{|\langle\chi_{b,2}({\rm 1P})|R|{\rm 2S}\rangle|^2}
{|\langle\chi_{b,1}({\rm 1P})|R|{\rm 2S}\rangle|^2}$ &
$\frac{|\langle\chi_{b,0}({\rm 1P})|R|{\rm 2S}\rangle|^2}
{|\langle\chi_{b,1}({\rm 1P})|R|{\rm 2S}\rangle|^2}$ \\
\hline
This Experiment    & $1.02\pm0.06\pm0.11$ & $0.75\pm0.09\pm0.26$\\
Previous World Average\cite{rev}& $0.92\pm0.11$ & $0.95\pm0.16$ \\
\end{tabular}
\end{table}

\section{Summary}
Based on the inclusive
measurement of photon energies taken from
$\Upsilon$(2S) data, we have measured the branching fractions
from the $\Upsilon$(2S) state to the $\chi_{b,J}$ triplet, as well
as the masses of the states in the triplet.
We find the ratio of mass splittings, 
$r$(1P)$\equiv(M_{J=2}-M_{J=1})/(M_{J=1}-M_{J=0})=0.54\pm0.02\pm0.02$,
substantially lower than the previous world average, but 
consistent with the expectation that $r$(1P)$<r$(2P).

\section{Acknowledgments}
We gratefully acknowledge the effort of the CESR staff in providing us with
excellent luminosity and running conditions.
This work was supported by 
the National Science Foundation,
the U.S. Department of Energy,
Research Corporation,
the Natural Sciences and Engineering Research Council of Canada, 
the A.P. Sloan Foundation, 
the Swiss National Science Foundation,
and the Alexander von Humboldt Stiftung.

\end{document}